\documentstyle[12pt]{article}
\def\be{\begin{equation}}
\def\ee{\end{equation}}

\begin{document}

\begin{center}
{\Large {\bf Review of Quantum Electromagnetodynamics}} \\

\vspace{0.5cm}
Rainer W. K\"uhne 
\end{center}

\vspace{1cm}

\noindent
{\bf Several years ago, I suggested a quantum field theory which has many 
attractive features. (1) It can explain the quantization of electric charge. 
(2) It describes symmetrized Maxwell equations. (3) It is manifestly 
covariant. (4) It describes local four-potentials. (5) It avoids the 
unphysical Dirac string. Here I will review the ideas which led to my 
model of magnetic monopoles including my prediction of the second kind 
of electromagnetic radiation. I will present also the mathematical formalism. 
Moreover I will suggest an experiment to verify the second kind 
of electromagnetic radiation and point out a possible observation of this 
radiation by August Kundt in 1885. Finally, I will list the many and 
far-reaching consequences, if this radiation will be confirmed by 
future experiments.}

\vspace{1cm}

\section{Introduction}
\noindent
The discovery of a second kind of light would  
be a multi-dimensional scientific revolution. It would shake the foundations 
of modern physics in many ways. It would be experimental 
evidence of physics beyond the standard theory of particle physics. 
The standard theory includes the Weinberg-Salam theory from 1967/1968 
and quantum chromodynamics from 1973. The observation would 
require not only that the theory of quantum electrodynamics formulated 
in 1948/1949 has to be extended. It would challenge also the 
Copenhagen interpretation of quantum mechanics formulated in 
1927/1928. Furthermore, the new kind of light would violate the relativity 
principle of special relativity from 1905 and would require a symmetrization 
of Maxwell's equations from 1873.

In the second chapter, I will review the ideas which led to my 
model of magnetic monopoles. The third chapter suggests an experiment 
to verify the second kind 
of electromagnetic radiation. In the fourth chapter, I will present 
the mathematical formalism. The fifth chapter deals with a possible 
observation of the second kind of electromagnetic radiation by August 
Kundt in 1885. In the sixth chapter, I will list the many and 
far-reaching consequences, if this radiation will be confirmed by 
future experiments.

\section{The Model}

The existence of the second kind of light was predicted theoretically. 
It can be understood by the following argumentation.

In 1948/1949 Tomonaga, Schwinger, and Feynman introduced quantum 
electrodynamics. It is the quantum field theory of electric and 
magnetic phenomena. This theory has one shortcoming. 
It cannot explain why electric charge is quantized, i.e. why it 
appears only in discrete units.

In 1931 Dirac \cite{Dirac} introduced the concept of magnetic monopoles. 
He has shown that any theory which includes magnetic monopoles 
requires the quantization of electric charge.

A theory of electric and magnetic phenomena which includes Dirac 
monopoles can be formulated in a manifestly covariant and 
symmetrical way if two four-potentials are used. Cabibbo and Ferrari 
in 1962 \cite{Cabibbo} were the first to formulate such a theory. It was 
examined in greater detail by later authors \cite{later}. 
Within the 
framework of a quantum field theory one four-potential corresponds 
to Einstein's electric photon from 1905 and the other four-potential 
corresponds to Salam's magnetic photon from 1966 \cite{Salam}.

In 1997 I have shown that the Lorentz force between an electric 
charge and a magnetic charge can be generated as follows \cite{1}. 
An electric charge 
couples via the well-known vector coupling with an electric photon and 
via a new type of tensor coupling, named velocity coupling, with a 
magnetic photon. This velocity coupling requires the existence of a 
velocity operator.

For scattering processes this velocity is the relative velocity 
between the electric charge and the magnetic charge just before 
the scattering. For emission and absorption processes there is no 
possibility of a relative velocity. The velocity is the absolute 
velocity of the electric charge just before the reaction.

The absolute velocity of a terrestrial laboratory was measured by 
the dipole anisotropy of the cosmic microwave background radiation. 
This radiation was detected in 1965 by Penzias and Wilson, its 
dipole anisotropy was detected in 1977 by Smoot, Gorenstein, and Muller 
\cite{Smoot}. 
The mean value of the laboratory's absolute velocity is 371 km/s. 
It has an annual sinusoidal period because of the Earth's motion 
around the Sun with 30 km/s. It has also a daily sinusoidal period 
because of the Earth's rotation with 0.5 km/s.

According to my model from 1997 \cite{1} each process that produces 
electric photons does create also magnetic photons. The cross-section 
of magnetic photons in a terrestrial laboratory is roughly one 
million times smaller than that of electric photons of the same energy. 
The exact value varies with time and has both the annual and the 
daily period.

As a consequence, magnetic photons are one million times harder to 
create, to shield, and to absorb than electric photons of the same 
energy.

The electric-magnetic duality is: 

\begin{center}
\begin{tabular}{lll}
electric charge & --- & magnetic charge \\
electric current & --- & magnetic current \\
electric conductivity & --- & magnetic conductivity \\
electric field strength & --- & magnetic field strength \\
electric four-potential & --- & magnetic four-potential \\
electric photon & --- & magnetic photon \\
electric field constant & --- & magnetic field constant \\
dielectricity number & --- & magnetic permeability
\end{tabular}
\end{center}

The refractive index of an insulator is the square root of the product of 
the dielectricity number and the magnetic permeability. Therefore it  
is invariant under a dual transformation. This means that electric and 
magnetic photon rays are reflected and refracted by insulators in the same 
way. Optical lenses cannot distinguish between electric and 
magnetic photon rays.

By contrast, electric and magnetic photon rays are reflected and refracted 
in a different way by metals. This is because electric conductivity and 
magnetic conductivity determine the reflection of light and they 
are not identical. The electric conductivity of a metal is several 
orders larger than the magnetic conductivity. 

\section{How to Verify the Magnetic Photon Rays}

The easiest test to verify/falsify the magnetic photon is to illuminate a 
metal foil of thickness $1,\ldots ,100\mu$m  by a laser beam (or any other 
bright light source) and to place a detector (avalanche diode or 
photomultiplier tube) behind the foil. If a single foil is used, then the 
expected reflection losses are less than 1\%. If a laser beam of the 
visible light is used, then the absorption losses are less than 15\%. My 
model \cite{1} predicts the detected intensity of the radiation to be 
\begin{equation}
f = r(v/c)^4
\end{equation}
times the intensity that would be detected 
if the metal foil were removed and the laser beam would directly illuminate 
the detector. Here
\begin{equation}
v = v_{sun} + v_{earth}\cos (2\pi t/T_e ) \cos ( \varphi_{ec}) 
+ v_{rotation} \cos(2\pi t/T_{rot}) \cos ( \varphi_{eq})
\end{equation}
is the absolute velocity of the laboratory. The absolute velocity 
of the Sun as measured by the dipole anisotropy of the cosmic microwave 
background radiation is
\begin{equation}
v_{sun} = (371 \pm 0.5) \mbox{km/s}.
\end{equation}
The mean velocity of the Earth around the Sun is
\begin{equation}
v_{earth} = 30 \mbox{km/s}.
\end{equation}
The rotation velocity of the Earth is
\begin{equation}
v_{rotation} = 0.5 \mbox{km/s} \cos ( \varphi ).
\end{equation}
The latitude of the dipole with respect to the ecliptic is
\begin{equation}
\varphi_{ec} = 15^{\circ}.
\end{equation}
The latitude of the dipole with respect to the equator (declination) is
\begin{equation}
\varphi_{eq} = 7^{\circ}.
\end{equation}
The latitude of the laboratory is
\begin{equation}
\varphi = 48^{\circ}
\end{equation}
for Strassbourg and Vienna and $\varphi = 43^{\circ}$ for Madison. 
The sidereal year is
\begin{equation}
T_e = 365.24 \mbox{days}.
\end{equation}
A sidereal day is
\begin{equation}
T_{rot} = 23\mbox{h}~ 56\mbox{min}.
\end{equation}
The zero point of the time, $t = 0$, is reached on December 9 at 0:00 local 
time. The speed of light is denoted by $c$. The factor for losses by 
reflection and absorption of magnetic photon rays of the visible light 
for a metal foil of thickness $1, \ldots ,100 \mu$m is
\begin{equation}
r = 0.8, \ldots , 1.0 .
\end{equation}
To conclude, my model \cite{1} predicts the value $f\sim 10^{-12}$. 
Two experiments have been tried to confirm this prediction. 
The first one was tried in Vienna/Austria in February 2002. 
The second one was done in Madison/Wisconsin in March 2002. 
Both experiments yielded the value $f\sim 10^{-15}$. 
The result is not yet conclusive as background effects such as stray light 
cannot yet be excluded with certainty. 
If the result turns out to be correct, then it has to be explained why 
it is roughly 1000 times smaller than my prediction. 

One possibility is that the prediction $f\sim 10^{-12}$ is strictly valid 
only for free charges. However, in condensed matter we have interactions 
of light with the electromagnetic field instead of interactions with 
free particles. In particle physics, too, we often do not have free 
charges. For example, the emission of synchrotron radiation occurs 
when the charged particles are within an external field. This means that 
the particles are off the mass shell. Here, too, the velocity coupling 
does not refer to the velocity of a charged particle. This is because 
the velocity of a particle which is not on the mass shell is not defined.  
The velocity coupling rather describes the velocity of the entire system, 
i.e. the centre of mass velocity of the particle and the field. 
Usually, this velocity is non-relativistic. 

Another possibility is that the magnetic photon model 
has to be modified. Other versions of the magnetic photon model were 
suggested by Singleton \cite{Singleton} (where the magnetic photon 
has nonzero rest mass) and Carneiro \cite{Carneiro} (where all 
interactions occur via vector coupling and where the photon propagator 
is more complicated).

\section{Formalism}
Let $J^{\mu}=(P, {\bf J})$ denote the electric four-current and 
$j^{\mu}=(\rho , {\bf j})$ the magnetic four-current. The 
well-known four-potential of the electric photon is 
$A^{\mu}=(\Phi , {\bf A})$.  The four-potential of the magnetic photon is 
$a^{\mu}=(\varphi , {\bf a})$. Expressed in three-vectors the symmetrized 
Maxwell equations read,
\begin{eqnarray}
\nabla\cdot {\bf E} & = & P \\
\nabla\cdot {\bf B} & = & \rho \\
\nabla\times {\bf E} & = & - {\bf j} - \partial_{t} {\bf B} \\
\nabla\times {\bf B} & = & + {\bf J} + \partial_{t} {\bf E}
\end{eqnarray}
and the relations between field strengths and potentials are
\begin{eqnarray}
{\bf E} & = & - \nabla\Phi - \partial_{t} {\bf A} -\nabla\times {\bf a} \\
{\bf B} & = & - \nabla\varphi - \partial_{t} {\bf a} +\nabla\times {\bf A}.
\end{eqnarray}
The Lagrangian for a spin 1/2 fermion field 
$\Psi $ of rest mass $m_{0}$, electric charge 
$Q$, and magnetic charge $q$ within an electromagnetic field can be 
constructed as follows. 
By using the tensors
\begin{eqnarray}
F^{\mu\nu} & \equiv & \partial^{\mu}A^{\nu}- \partial^{\nu}A^{\mu} \\
f^{\mu\nu} & \equiv & \partial^{\mu}a^{\nu}- \partial^{\nu}a^{\mu}
\end{eqnarray}
the Lagrangian of the Dirac fermion within the electromagnetic field reads, 
\begin{eqnarray}
{\cal L} & = & - \frac{1}{4}F_{\mu\nu}F^{\mu\nu}
               - \frac{1}{4}f_{\mu\nu}f^{\mu\nu} 
  + \bar\Psi i\gamma^{\mu}\partial_{\mu}\Psi - m_{0}\bar\Psi \Psi 
 \nonumber \\
 & &  -Q\bar\Psi \gamma^{\mu}\Psi A_{\mu} - q\bar\Psi \gamma^{\mu}\Psi a_{\mu}
   +Q\bar\Psi \gamma^{5}\sigma^{\mu\nu}u_{\nu}\Psi a_{\mu} 
 +q\bar\Psi \gamma^{5}\sigma^{\mu\nu}u_{\nu}\Psi A_{\mu}.
\end{eqnarray}
By using the Euler-Lagrange equations we obtain the Dirac equation
\begin{equation}
(i\gamma^{\mu}\partial_{\mu}-m_{0})\Psi = (Q\gamma^{\mu}A_{\mu}
+q\gamma^{\mu}a_{\mu} 
 -Q\gamma^{5}\sigma^{\mu\nu}u_{\nu}a_{\mu} 
 -q\gamma^{5}\sigma^{\mu\nu}u_{\nu}A_{\mu})\Psi .
\end{equation}
By introducing the four-currents
\begin{eqnarray}
J^{\mu} & = & Q\bar\Psi \gamma^{\mu}\Psi -q\bar\Psi \gamma^{5}\sigma^{\mu\nu}
u_{\nu}\Psi \\
j^{\mu} & = & q\bar\Psi \gamma^{\mu}\Psi -Q\bar\Psi \gamma^{5}\sigma^{\mu\nu}
u_{\nu}\Psi
\end{eqnarray}
the Euler-Lagrange equations yield the two Maxwell equations
\begin{eqnarray}
J^{\mu} & = & \partial_{\nu}F^{\nu\mu} = \partial^{2}A^{\mu} 
- \partial^{\mu}\partial^{\nu}A_{\nu} \\
j^{\mu} & = & \partial_{\nu}f^{\nu\mu} = \partial^{2}a^{\mu} 
- \partial^{\mu}\partial^{\nu}a_{\nu}.
\end{eqnarray}
Evidently, the two Maxwell equations are invariant under the 
$U(1)\times U'(1)$ gauge transformations
\begin{eqnarray}
A^{\mu} & \rightarrow & A^{\mu}-\partial^{\mu}\Lambda \\
a^{\mu} & \rightarrow & a^{\mu}-\partial^{\mu}\lambda .
\end{eqnarray}
Furthermore, the four-currents satisfy the continuity equations
\begin{equation}
0=\partial_{\mu}J^{\mu}= \partial_{\mu}j^{\mu}.
\end{equation}
The electric and magnetic field are related to the tensors above by
\begin{eqnarray}
E^{i} & = & F^{i0}- \frac{1}{2}\varepsilon^{ijk}f_{jk} \\
B^{i} & = & f^{i0}+ \frac{1}{2}\varepsilon^{ijk}F_{jk}.
\end{eqnarray}
Finally, the Lorentz force is
\begin{equation}
K^{\mu}  = Q(F^{\mu\nu}+ \frac{1}{2}\varepsilon^{\mu\nu\varrho\sigma}
              f_{\varrho\sigma})u_{\nu} 
 + q(f^{\mu\nu}- \frac{1}{2}\varepsilon^{\mu\nu\varrho\sigma}
              F_{\varrho\sigma})u_{\nu},
\end{equation}
where $\varepsilon^{\mu\nu\varrho\sigma}$ denotes the totally 
antisymmetric tensor.

\section{Possible Observation of Magnetic Photon Rays}

\noindent
In Strassbourg in 1885, August Kundt \cite{Kundt} passed sunlight through 
red glass, a polarizing 
Nicol, and platinized glass which was covered by an iron layer. The entire 
experimental setup was placed within a magnetic field. With the naked eye, 
Kundt measured the Faraday rotation of the polarization plane generated by 
the transmission of the sunlight through the iron layer. His result was a 
constant maximum rotation of the polarization plane per length of 
$418,000^{\circ}$/cm or $1^{\circ}$ per 23.9nm. He verified this result 
until thicknesses of up to 210nm and rotations of up to $9^{\circ}$. 

In one case, on a very clear day, he observed the penetrating sunlight for 
rotations of up to $12^{\circ}$. Unfortunately, he has not given the 
thickness of this particular iron layer he used. But if his result of a 
constant maximum rotation per length can be applied, then the corresponding 
layer thickness was $\sim 290$nm.

Let us recapitulate some classical electrodynamics to determine the 
behavior of light within iron. The penetration depth of light in a 
conductor is
\be
\delta = \frac{\lambda}{2\pi\gamma},
\ee
where the wavelength in vacuum can be expressed by its frequency 
according to $\lambda = 1/ \sqrt{\nu^2 \varepsilon_0 \mu_0}$. The 
extinction coefficient is
\be
\gamma = \frac{n}{\sqrt{2}}\left[ -1 + \left( 1+ \left( 
\frac{\sigma}{2\pi\nu\varepsilon_0\varepsilon_r} \right)^2 \right) ^{1/2} 
\right] ^{1/2} ,
\ee
where the refractive index is $n=\sqrt{\varepsilon_r \mu_r }$. For 
metals we get the very good approximation
\be
\delta\approx\left( \frac{1}{\pi\mu_0\mu_r\sigma\nu} \right) ^{1/2}.
\ee
The specific resistance of iron is
\be
1/ \sigma = 8.7\times 10^{-8}\Omega\mbox{m},
\ee
its permeability is $\mu_r \geq 1$. For red light of $\lambda =630$nm 
and $\nu =4.8\times 10^{14}$Hz we get the penetration depth
\be
\delta = 6.9\mbox{nm}.
\ee

Only a small fraction of the sunlight can enter the iron layer. Three 
effects have to be considered. (i) The red glass allows the penetration 
of about $\varepsilon_1 \sim 50\% $ of the sunlight only. (ii) Only 
$\varepsilon_2 =2/ \pi \simeq 64\% $ of 
the sunlight can penetrate the polarization filter. (iii) Reflection 
losses at the surface of the iron layer have to be considered. The 
refractive index for electric photon light is given by
\begin{equation}
\bar n^{2} = \frac{n^{2}}{2} \left( 1+ \sqrt{ 1+ \left( 
\frac{\sigma}{2\pi\varepsilon_0 \varepsilon_r \nu} \right)^{2}} \right).
\end{equation}
For metals we get the very good approximation
\begin{equation}
\bar n \simeq \sqrt{ \frac{\mu_r \sigma}{4\pi\varepsilon_0 \nu}}.
\end{equation}
The fraction of the sunlight which is not reflected is
\begin{equation}
\varepsilon_3 = \frac{2}{1+ \bar n}= 
\frac{2}{1+ \sqrt{\mu_r \sigma /(4\pi\varepsilon_0 \nu )}}
\end{equation}
and therefore $\varepsilon_3 \simeq 0.13$ for the system considered. Taken 
together, 
the three effects allow only 
$\varepsilon_1 \varepsilon_2 \varepsilon_3 \sim 4\% $ 
of the sunlight to enter the iron layer. 

The detection limit of the naked eye is $10^{-13}$ times the brightness 
of sunlight provided the light source is pointlike. For an extended 
source the detection limit depends on the integral and the surface 
brightness. The detection limit for a source as extended as the Sun 
(0.5$^{\circ}$ diameter) is $l_d \sim 10^{-12}$ times the brightness of 
sunlight. If 
sunlight is passed through an iron layer (or foil, respectively), then it  
is detectable with the naked eye only if it has passed not more than 
\be
( \ln (1/l_d ) + \ln ( \varepsilon_1 \varepsilon_2 \varepsilon_3 )) \delta 
\sim 170 \mbox{nm}. 
\ee
Reflection losses by haze in the atmosphere further reduce this value. 

Kundt's observation can hardly be explained with classical electrodynamics. 
Air bubbles within the metal layers cannot explain Kundt's observation, 
because air does not generate such a large rotation. Impurities, such 
as glass, which do generate an additional rotation, cannot completely be 
ruled out as the explanation. However, impurities are not a likely 
explanation, because Kundt was able to reproduce his observation by using 
several layers which he examined at various places. 

Quantum effects cannot explain the observation, because they decrease 
the penetration depth, whereas an increment would be required.

The observation may become understandable if Kundt has observed a 
second kind of electromagnetic radiation, the magnetic photon rays. 
I predict their penetration depth to be 
\be
\delta_m = \delta (c/v)^2 \sim 5\mbox{mm}.
\ee
To learn whether Kundt has indeed observed magnetic photon rays, his 
experiment has to be repeated.

\section{Consequences}

The observation of magnetic photon rays would be a multi-dimensional 
revolution in physics. Its implications would be far-reaching.

(1) The experiment would provide evidence of a second kind of electromagnetic 
radiation. The penetration depth of these magnetic photon rays is 
roughly one million times greater than that of
electric photon light of the same wavelength. Hence, these new rays may find 
applications in medicine where X-ray and ultrasonic diagnostics are 
not useful. X-ray examinations include a high risk of radiation damages, 
because the examination of teeth requires high intensities of 
X-rays and genitals are too sensible to radiation damages. Examinations 
of bones and the brain may also become possible.

(2) The experiment would confirm the existence of a new vector gauge boson, 
Salam's magnetic photon from 1966. It has the same quantum numbers as 
Einstein's electric photon, i.e. spin of one, negative parity, zero 
rest mass, and zero charge.

(3) A positive result would provide evidence of an extension of quantum 
electrodynamics which includes a symmetrization of Maxwell's 
equations from 1873.

(4) The experiment would provide indirect evidence of Dirac's magnetic 
monopoles from 1931 and the explanation of the quantization of electric 
charge. This quantization is known since Rutherford's discovery of the 
proton in 1919.

(5) My model describes both an electric current and a magnetic current, 
even in experimental situations which do not include magnetic charges. 
This new magnetic current has a larger specific resistance in conductors 
than the electric current. It may find applications in electronics.

(6) Dirac noticed in 1931 that the coupling constant of magnetic 
monopoles is much greater than unity. This raises new questions 
concerning the perturbation theory, the renormalizability, and the 
unitarity of quantum field theories.

(7) The intensity of the magnetic photon rays should depend on 
the absolute velocity of the laboratory. The existence of the 
absolute velocity would violate Einstein's relativity principle of special 
relativity from 1905. It would be interesting to learn whether 
there exist further effects of absolute motion.

(8) The supposed non-existence of an absolute rest frame was the only 
argument against the existence of a luminiferous aether. If the 
absolute velocity does exist, we have to ask whether aether 
exists and what its nature is.

(9) When in 1925 Heisenberg introduced quantum mechanics, he argued 
that motion does not exist in this theory. This view is taken also 
in the Copenhagen interpretation of quantum mechanics formulated in 
1927/1928 by Heisenberg and Bohr. The appearance of a velocity 
operator in my model challenges this Copenhagen interpretation. 
Mathematically, the introduction of a velocity (and force) operator 
means that quantum mechanics has to be described not only by partial but 
also by ordinary differential equations.

(10) Magnetic photon rays may contribute to our understanding of 
several astrophysical and high energy particle physics phenomena 
where relativistic absolute velocities appear and where electric 
and magnetic photon rays are expected to be created in comparable 
intensities.

(11) Finally, the other interactions may show similar dualities. 
The new dual partners of the known gauge bosons would be the 
magnetic photon, the isomagnetic W- and Z-boson, and the 
chromomagnetic gluons. In 1999 I argued that the dual 
partner of the graviton would be the tordion \cite{2}. This boson has a spin 
of three and is required by Cartan's torsion theory from 1922 which 
is an extension of Einstein's general relativity from 1915.

\end{document}